\pdfoutput=1

\documentclass[11pt,a4paper]{article}
\usepackage{bbm}
\usepackage{bm}
\usepackage{graphicx}
\usepackage{jheppub}
\usepackage{mathrsfs}

\let\de=\partial
\DeclareMathOperator{\re}{Re}% real part
\DeclareMathOperator{\tr}{Tr}% trace
\DeclareMathOperator{\cn}{cn}% Jacobi's elliptic function
\DeclareMathOperator{\dn}{dn}% Jacobi's elliptic function
\DeclareMathOperator{\sn}{sn}% Jacobi's elliptic function
\newcommand{\imag}{\text{i}}% imaginary unit
\newcommand{\En}{\mathscr{E}}% energy of the soliton
\newcommand{\Ha}{\mathscr{H}}% Hamiltonian symbol
\newcommand{\La}{\mathscr{L}}% Lagrangian symbol
\newcommand\openone{\mathbbm{1}}% unit matrix
\newcommand{\dd}{\text{d}}% Roman d for differentials
\newcommand{\he}[1]{#1^\dagger}% Hermitian conjugate
\newcommand{\vek}[1]{\bm{#1}}% spatial vector
\newcommand{\ave}[1]{\langle{#1}\rangle}% vacuum expectation value

\title{Chiral soliton lattice and charged pion condensation\\
in strong magnetic fields}

\author[a]{Tom\'a\v{s} Brauner}
\author[b]{and Naoki Yamamoto}
\affiliation[a]{Faculty of Science and Technology, University of Stavanger,\\
N-4036 Stavanger, Norway}
\affiliation[b]{Department of Physics, Keio University,\\
Yokohama 223-8522, Japan}
\emailAdd{tomas.brauner@uis.no}
\emailAdd{nyama@rk.phys.keio.ac.jp}

\abstract{The Chiral Soliton Lattice (CSL) is a state with a periodic array of topological solitons that spontaneously breaks parity and translational symmetries. Such a state is known to appear in chiral magnets. We show that CSL also appears as a ground state of quantum chromodynamics at nonzero chemical potential in a magnetic field. By analyzing the fluctuations of the CSL, we furthermore demonstrate that in strong but achievable magnetic fields, charged pions undergo Bose-Einstein condensation. Our results, based on a systematic low-energy effective theory, are model-independent and fully analytic.}

\keywords{Phase Diagram of QCD, Effective Field Theories, Topological States of Matter}

\begin{document}

\maketitle

%%%%%%%%%%%%%%%%%%%%%%%%%%%%%%%%%%%%%%%%%%%%%%%%%%%%%%%%%%%%%%%%%%%%

\section{Introduction}

The Chiral Soliton Lattice (CSL) is a periodic, parity-violating topological soliton. CSL-type structures have been studied in condensed-matter systems such as chiral magnets~\cite{Dzyaloshinsky:1964dz} and cholesteric liquid crystals~\cite{Gennes}. In this paper, we show that CSL appears as the ground state of Quantum ChromoDynamics (QCD) at nonzero baryon chemical potential in an external magnetic field. The existence of a ``stack of parallel domain walls'' under such conditions was previously conjectured in ref.~\cite{Son:2007ny}. Here we provide an \emph{exact} solution to the underlying equations of motion, and thus firmly establish the presence of CSL in the QCD phase diagram.

Our argument is based on a systematic, model-independent low-energy effective theory and is thus under theoretical control; this is a prediction for QCD \emph{per se}. Our topological soliton lattice is chiral in that it breaks parity spontaneously.\footnote{Note that in condensed-matter physics, the term ``chiral symmetry'' usually refers to a discrete parity symmetry, which is different from the continuous chiral symmetry $\text{SU}(N_\text{f})_\text{L}\times\text{SU}(N_\text{f})_\text{R}$ of QCD, where $N_\text{f}$ is the number of quark flavors.} To the best of our knowledge, this is the first realization of CSL in the context of high-energy physics. Our results also provide a new example of hierarchical symmetry breaking~\cite{Raby:1979my,Nambu:1998vi}: chiral symmetry of QCD is first spontaneously broken, giving rise to pions as the pseudo-Nambu-Goldstone bosons. Subsequently, pion dynamics in strong magnetic fields generates the CSL with broken continuous translational symmetry, leading to a phonon as the Nambu-Goldstone boson. 

Throughout the text, we use the natural units in which the Planck's constant $\hbar$, the speed of light $c$ and the elementary electric charge $e$ are all set to one.

%%%%%%%%%%%%%%%%%%%%%%%%%%%%%%%%%%%%%%%%%%%%%%%%%%%%%%%%%%%%%%%%%%%%

\section{Low-energy effective theory}

The low-energy dynamics of QCD is dominated by the spontaneously broken chiral symmetry, and can thus be described by an effective theory for the ensuing Nambu-Goldstone bosons (pions): the chiral perturbation theory. The predictions of this effective theory are organized by a derivative expansion, controlled by the parameter $p/(4\pi f_\pi)$, where $p$ is a characteristic momentum and $f_\pi$ is the pion decay constant~\cite{Weinberg:1978kz,Manohar:1983md}. The leading order of the chiral Lagrangian for two quark flavors reads~\cite{Gasser:1983yg}
\begin{equation}
\La=\frac{f_\pi^2}4\left[\tr(D_\mu\he\Sigma D^\mu\Sigma)+2m_\pi^2\re\tr\Sigma\right].
\label{lagrangian}
\end{equation}
Here $\Sigma$ is a $2\times2$ unitary matrix containing the pion degrees of freedom. It couples to an external electromagnetic gauge field $A_\mu$ via the covariant derivative $D_\mu\Sigma\equiv\de_\mu\Sigma-\imag[Q_\mu,\Sigma]$, where $Q_\mu\equiv A_\mu\frac{\tau_3}2$, and $\vec\tau$ denotes the set of Pauli matrices.

The QCD vacuum corresponds to $\ave\Sigma=\openone$. However, in presence of a nonzero baryon chemical potential $\mu$ and an external magnetic field $\vek B$, a nontrivial neutral pion structure may appear~\cite{Son:2007ny}. Setting thus $\Sigma=e^{\imag\tau_3\phi}$, the anticipated time-independent neutral pion background $\phi$ is obtained by minimization of the energy functional. The latter in turn follows from the Hamiltonian density that equals, up to a constant,\footnote{The offset $m_\pi^2f_\pi^2$ is chosen so that the energy density vanishes in the QCD vacuum.}
\begin{equation}
\Ha=\frac{f_\pi^2}2(\vek\nabla\phi)^2+m_\pi^2f_\pi^2(1-\cos\phi)-\frac{\mu}{4\pi^2}\vek B\cdot\vek\nabla\phi.
\label{hamiltonian}
\end{equation}
The first two terms follow from eq.~\eqref{lagrangian}. The last term, on the other hand, is a topological surface term that arises from the anomalous coupling of neutral pions to the electromagnetic field~\cite{Son:2004tq,Son:2007ny}. Without loss of generality, we will orient the uniform external magnetic field along the $z$-axis, $\vek B\equiv(0,0,B)$. It is then easy to see that in the chiral limit ($m_{\pi}=0$), the total energy is minimized when
\begin{equation}
\phi(z)=\frac{\mu Bz}{4\pi^2f_\pi^2}.
\label{CSL_chiral_limit}
\end{equation}
This solution is characterized by a new, emergent momentum scale, $p_\text{CSL}\equiv\mu B/(4\pi^2f_\pi^2)$. Validity of the derivative expansion of the effective theory then requires that $p_\text{CSL}\ll4\pi f_\pi$. The same condition ensures that the non-anomalous next-to-leading order contributions to the chiral Lagrangian, containing four derivatives of $\phi$, are negligible compared to terms included in eq.~\eqref{hamiltonian}. Unless explicitly stated otherwise, the results presented in this paper are therefore exact up to a relative correction of the order $p_\text{CSL}^2/(4\pi f_\pi)^2$. Note that the scale $p_\text{CSL}$ can be arbitrarily small, thus creating a hierarchy of scales for symmetry breaking.

%%%%%%%%%%%%%%%%%%%%%%%%%%%%%%%%%%%%%%%%%%%%%%%%%%%%%%%%%%%%%%%%%%%%

\section{Chiral soliton lattice}

For nonzero $m_\pi$, one needs to take into account the competition between the periodic potential and the anomalous term in eq.~\eqref{hamiltonian}. This competition leads to the CSL solution, as we will now see (see also ref.~\cite{Kishine20151} for a similar calculation in a different context). Since the derivatives of $\phi$ with respect to $x,y$ only enter the Hamiltonian~\eqref{hamiltonian} through the first term, the ground state necessarily features a modulation in the $z$-direction only. The equation of motion for such a one-dimensional configuration $\phi(z)$ then reads
\begin{equation}
\de_z^2\phi=m_{\pi}^2\sin\phi.
\label{CSL_Eom}
\end{equation}
Note that this is mathematically equivalent to the equation of motion of a simple pendulum, which can be solved in a closed form in terms of the Jacobi elliptic functions. We thus get
\begin{equation}
\cos\frac{\phi(\bar z)}2=\sn(\bar z,k), 
\label{CSL_solution}
\end{equation}
where $\bar z\equiv zm_\pi/k$ is a dimensionless coordinate and $k$ ($0 \leq k \leq 1$) is the elliptic modulus.

Let us look at the properties of this solution. First, observe that it has a lattice structure. For $(2n-1)K \leq \bar z \leq (2n+1)K$ with $n$ integer and $K\equiv K(k)$ the complete elliptic integral of the first kind, the angle $\phi$ varies from $2(n-1)\pi$ to $2n\pi$. Hence, the period of the lattice is given by
\begin{equation}
\ell=\frac{2kK(k)}{m_{\pi}}.
\end{equation}
Second, the anomalous term in eq.~\eqref{hamiltonian} gives rise to a local baryon charge and magnetization~\cite{Son:2007ny},
\begin{equation}
n_\text{B}(z)=\frac{B}{4\pi^2}\de_z \phi(z),\qquad
m(z)=\frac{\mu}{4\pi^2}\de_z \phi(z).
\label{topcharge}
\end{equation}
Consequently, each unit cell of the lattice carries a baryon charge and magnetic moment per unit area in the $xy$ plane,
\begin{equation}
\frac{N_\text{B}}{S}=\frac{B}{2\pi},\qquad
\frac{M}{S}=\frac{\mu}{2\pi}.
\label{N_B}
\end{equation}
Both of these are topological charges in that they depend only on the values at the boundary of a unit cell such as $\phi(-\ell/2)=-2\pi$ and $\phi(\ell/2)=0$, but not on details of the function $\phi(z)$. 

\begin{figure}
\begin{center}
\includegraphics[width=0.75\textwidth]{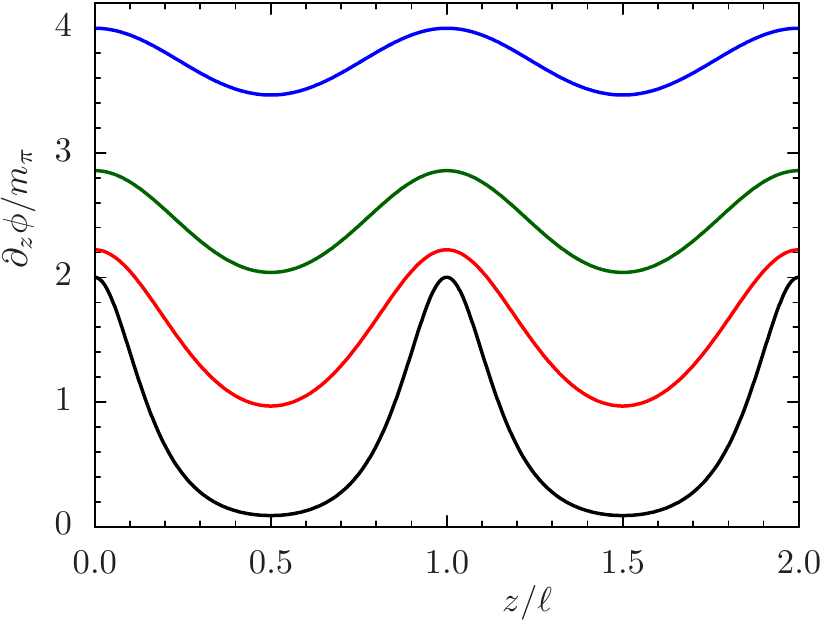}
\end{center}
\caption{The gradient of $\phi$, proportional to the topological charge densities~\eqref{topcharge}, as a function of position. The various curves correspond to $k$ of $0.999$ (black), $0.9$ (red), $0.7$ (green), and $0.5$ (blue).}
\label{fig:CSL}
\end{figure}

We conclude that the solution~\eqref{CSL_solution} exhibits a periodic array of topological solitons carrying baryon charge and magnetic moment. Being a coordinate-dependent condensate of pseudoscalar mesons, the solution breaks parity. It may be called CSL in analogy to the structure found in chiral magnets~\cite{Kishine}. The profile of the CSL, as given by the gradient of the phase $\phi$, is displayed in figure~\ref{fig:CSL} for several different values of $k$.

%%%%%%%%%%%%%%%%%%%%%%%%%%%%%%%%%%%%%%%%%%%%%%%%%%%%%%%%%%%%%%%%%%%%

\section{Ground state}
\label{sec:ground_state}

So far we treated $k$ as a free parameter. Its value in the ground state is determined by minimization of the total Hamiltonian. Using elementary properties of the Jacobi elliptic functions~\cite{Abramowitz:1972}, one readily shows that the energy of each soliton with period $\ell$ per unit area in the $xy$ plane is
\begin{equation}
\frac\En S = \frac{\En_\text{norm}}S-\frac{\mu B}{2\pi}.
\label{E_soliton}
\end{equation}
Here the first term stems from the non-anomalous part of the Hamiltonian~\eqref{hamiltonian},
\begin{equation}
\frac{\En_\text{norm}}S=4m_\pi f_\pi^2\left[\frac{2E(k)}{k}+\left(k- \frac{1}{k}\right)K(k)\right]\equiv F(k),
\label{E_norm}
\end{equation}
where we used the identity for the complete elliptic integral of the second kind, $E(k)$,
\begin{equation}
E(k)=\int_0^K\dd\bar z\,\dn^2(\bar z,k).
\label{elliptic2}
\end{equation}
Note that eq.~\eqref{E_soliton} can also be expressed as $({\cal E}_\text{norm}-\mu N_\text{B})/S$. This is in accord with the usual expression for the Hamiltonian at finite baryon density, $\Ha=\Ha_\text{norm}-\mu n_\text{B}$.

\begin{figure}
\begin{center}
\includegraphics[width=0.75\textwidth]{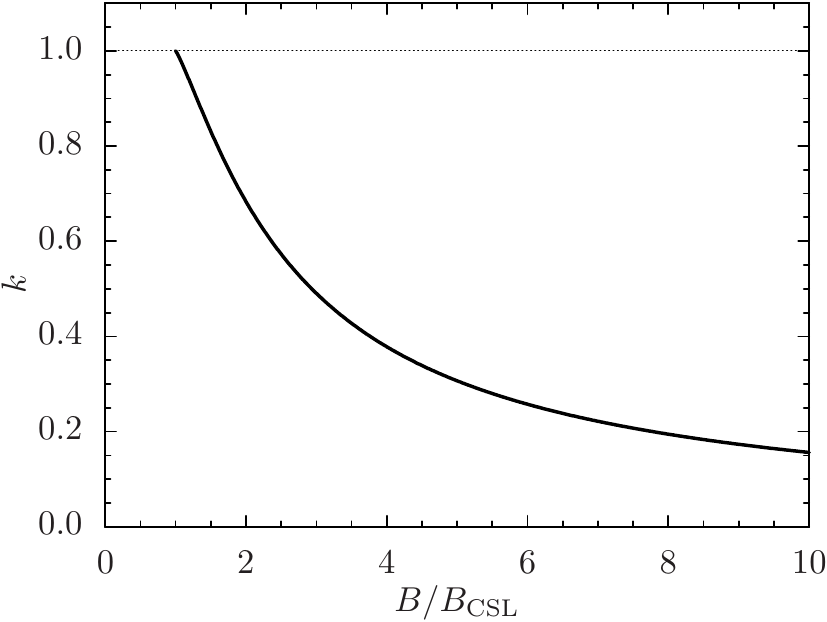}
\end{center}
\caption{The value of the elliptic modulus $k$ in the ground state as a function of magnetic field.}
\label{fig:k}
\end{figure}
In the limit $k\to1$, the Jacobi elliptic functions reduce to the hyperbolic functions and $\ell\rightarrow \infty$, which corresponds to a single pion domain wall, considered in ref.~\cite{Son:2007ny}. In this case, the energy density becomes $F(1)=8m_\pi f_\pi^2$, which indeed coincides with that of the pion domain wall. In the general CSL case, the total energy of the system of length $L$ in the $z$-direction is given by
\begin{equation}
\En_\text{tot}=\frac{L}{\ell}\En=\frac{V m_{\pi}}{2k K(k)}\left[F(k)-\frac{\mu B}{2\pi}\right],
\end{equation}
where $V\equiv LS$ is the volume of the system. Minimization of the total energy $\En_\text{tot}$ at fixed volume $V$ with respect to $k$ leads to the simple condition,
\begin{equation}
\frac{E(k)}k=\frac{\mu B}{16\pi m_\pi f_\pi^2}.
\label{condition}
\end{equation}
This condition fixes the optimal $k$ for given external parameters, $\mu$ and $B$. Since the left-hand side of eq.~\eqref{condition} is bounded from below as $E(k)/k\geq1$ ($0 \leq k \leq 1$), the CSL solution exists if and only if the following condition is satisfied,
\begin{equation}
\mu B>\mu B_\text{CSL}\equiv16\pi m_\pi f_\pi^2.
\label{main}
\end{equation}
\begin{figure}
\begin{center}
\includegraphics[width=0.75\textwidth]{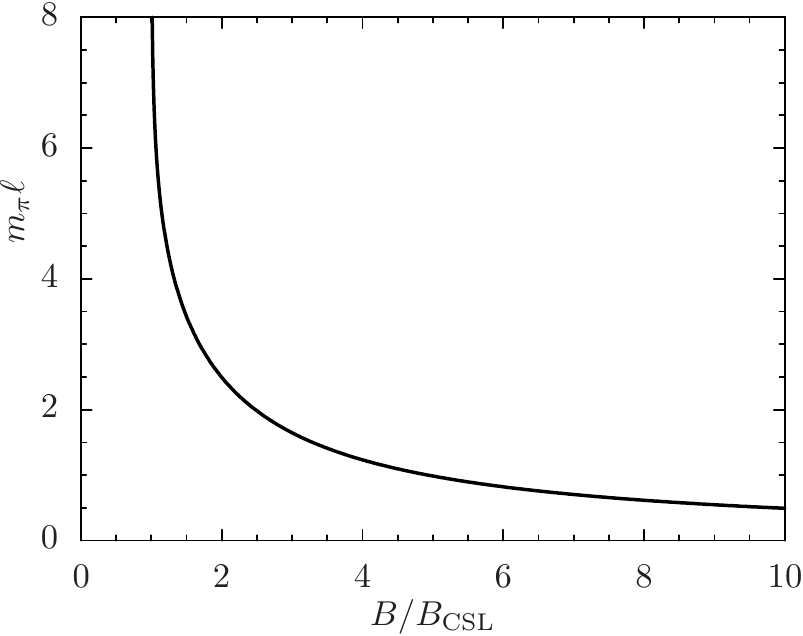}
\end{center}
\caption{The period of the CSL $\ell$ in units of inverse pion mass as a function of magnetic field.}
\label{fig:ell}
\end{figure}%
This is one of our main results. The value of $k$ in the ground state for magnetic fields above $B_\text{CSL}$ is displayed in figure~\ref{fig:k}. By combining this with figure~\ref{fig:CSL}, we can see that very close to $B_\text{CSL}$, the CSL takes the form of a periodic array of thin domain walls, but with increasing magnetic field it turns into a smooth, cosine-like profile. To complete the picture, we also show in figure~\ref{fig:ell} the period of the CSL in units of the inverse pion mass as a function of the magnetic field. This indicates large separation of the domain walls just above the critical field $B_\text{CSL}$. As a result of using the inverse pion mass as a unit for the lattice spacing, the chiral limit has to be treated with some care. In this case, the CSL profile is given by eq.~\eqref{CSL_chiral_limit} and the period is determined as the distance over which $\phi$ increases by $2\pi$,
\begin{equation}
\lim_{m_\pi\to0}\ell=\frac{8\pi^3f_\pi^2}{\mu B}.
\end{equation}

One can show that when the condition~\eqref{main} is satisfied, CSL is energetically more favorable than the usual QCD vacuum for $\mu \leq m_\text{N}$, and than nuclear matter for $\mu \approx m_\text{N}$, where $m_\text{N}$ is the nucleon mass. To see this, just combine eqs.~\eqref{E_norm} and~\eqref{condition} to write the CSL energy per unit area as
\begin{equation}
\frac\En S=4m_\pi f_\pi^2\left(k-\frac1k\right)K(k)<0,
\end{equation}
where we used that $0\leq k<1$ and $K(k)>0$. When $\mu\leq m_\text{N}$, the CSL energy is therefore always lower than that of the usual QCD vacuum independently of $\mu$ and $B$. Likewise, eqs.~\eqref{E_norm} and~\eqref{condition} give us the CSL energy per unit baryon number,
\begin{equation}
\frac{\En_\text{norm}}{N_\text{B}}=\mu+\frac{8\pi m_\pi f_\pi^2}B\left(k-\frac1k\right)K(k),
\end{equation}
which analogously implies that for $\mu\approx m_\text{N}$, $\En_\text{norm}/N_\text{B}\leq m_\text{N}$, and so CSL is energetically favorable compared with nuclear matter as well. Therefore, eq.~\eqref{main} is the necessary and sufficient condition for the realization of CSL within the range of validity of the low-energy effective theory~\eqref{lagrangian}.

%%%%%%%%%%%%%%%%%%%%%%%%%%%%%%%%%%%%%%%%%%%%%%%%%%%%%%%%%%%%%%%%%%%%

\section{Phonons}

The CSL configuration spontaneously breaks continuous translational symmetry in the $z$-direction down to a discrete one, and continuous rotations down to rotations around the $z$-axis. This is a new example of hierarchical symmetry breaking~\cite{Raby:1979my,Nambu:1998vi} in QCD. Based on the general properties of broken spacetime symmetries~\cite{Low:2001bw,Watanabe:2013iia}, we expect \emph{one} Nambu-Goldstone boson: the phonon of the soliton lattice.

\begin{figure}
\begin{center}
\includegraphics[width=0.75\textwidth]{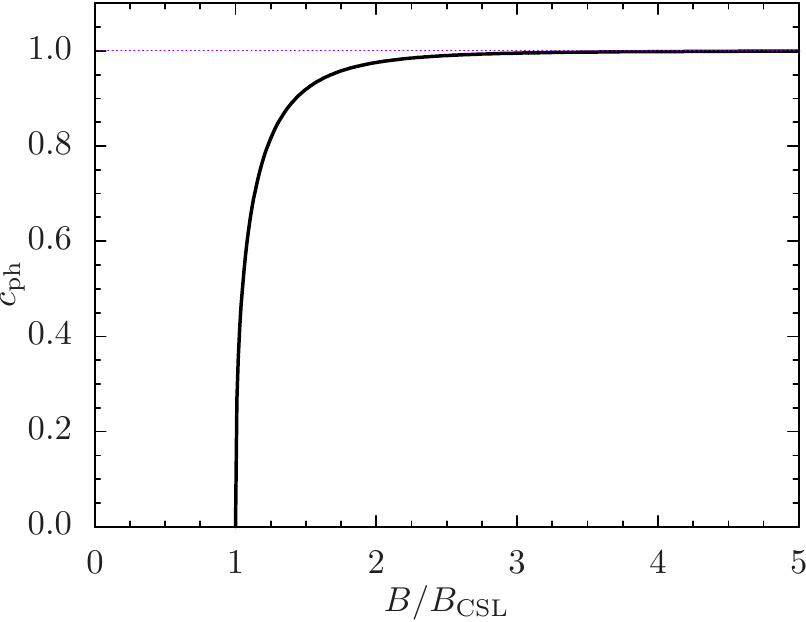}
\end{center}
\caption{Group velocity of the CSL phonon as a function of magnetic field.}
\label{fig:cph}
\end{figure}

The phonon spectrum is most easily found by analyzing the linear perturbations of the CSL solution using eq.~\eqref{CSL_Eom}. Restoring full spacetime dependence of the fluctuation $\pi(x)$, its linearized equation of motion reads
\begin{equation}
(\Box+m_\pi^2\cos\phi)\pi=0.
\end{equation}
Using eq.~\eqref{CSL_solution} and Fourier transforming in all spacetime coordinates but $z$, this is seen to be equivalent to
\begin{equation}
\left[-\de_{\bar z}^2+2k^2\sn^2(\bar z,k)\right]\pi=\frac{k^2}{m_\pi^2}[m_\pi^2+\omega^2-(p_x^2+p_y^2)]\pi,
\label{eomaux}
\end{equation}
where $p_{x,y}$ and $\omega$ are the phonon momentum and frequency. Now the left-hand side is the Lam\'e operator with $n=1$. The Bloch-type eigenstates of this operator are well-known in the literature~\cite{Whittaker:1927,Sutherland:1973zz}, here we focus just on the spectrum though. For $n=1$, this consists of two bands. The eigenvalues of the first band are bounded from below by $k^2$, which ensures that the phonon is gapless~\cite{Lame:Finkel,Lame:Iachello}. The phonon group velocity, defined by $c_\text{ph}\equiv{\dd\omega}/{\dd p_z}$, can be obtained from the dispersion relation, presented in ref.~\cite{Lame:Iachello},
\begin{equation}
c_\text{ph}=\sqrt{1-k^2}\frac{K(k)}{E(k)},
\label{phonondisprel}
\end{equation}
where $k$ is given implicitly by eq.~\eqref{condition}. (An alternative derivation of this result is provided in appendix~\ref{app:phonon}.) The group velocity vanishes at the critical magnetic field for CSL formation $B_\text{CSL}$, given by eq.~\eqref{main}, where $k=1$, and rapidly approaches the speed of light as the magnetic field increases, see figure~\ref{fig:cph}. The full dispersion relation of the phonon reads
\begin{equation}
\omega^2=p_x^2+p_y^2+(1-k^2)\left[\frac{K(k)}{E(k)}\right]^2p_z^2+\mathcal O(p_z^4),
\label{phonondisp}
\end{equation}
where $p_z$ is the phonon crystal (Bloch) momentum measured from the bottom of the valence band.

%%%%%%%%%%%%%%%%%%%%%%%%%%%%%%%%%%%%%%%%%%%%%%%%%%%%%%%%%%%%%%%%%%%%

\section{Charged pions}

The phonon represents the neutral pion fluctuation of the CSL state. To analyze the \emph{charged} pion fluctuations, we need to expand the full Lagrangian~\eqref{lagrangian} around the CSL solution, see appendix~\ref{app:EFT} for details. It is convenient to preserve unitarity of the matrix field variable by setting $\Sigma=\Sigma_0 U$, where $\Sigma_0\equiv e^{\imag\tau_3\phi}$. The unitary matrix $U$ can then be expanded in the pion fields $\vec\pi$ as usual. The bilinear part of the Lagrangian reads
\begin{equation}
\La_\text{bilin}=\frac12(\de_\mu\vec\pi)^2+(A^\mu-\de^\mu\phi)(\pi_1\de_\mu\pi_2-\pi_2\de_\mu\pi_1)+\frac12 A^\mu A_\mu(\pi_1^2+\pi_2^2)-\frac12m_\pi^2\vec\pi^2\cos\phi.
\label{Lbilin}
\end{equation}
Next we fix the gauge for the electromagnetic vector potential as $\vek A=(0,Bx,0)$ and introduce the charge eigenstates $\pi^\pm\equiv(\pi_1\pm\imag\pi_2)/\sqrt2$. Upon Fourier transforming in time and the $y$-coordinate, the equation of motion for $\pi^+$ takes the form
\begin{equation}
\omega^2\pi^+=\left[-\de_x^2+B^2\left(x-\frac{p_y}B\right)^2\right]\pi^++(-\de_z^2+2\imag\phi'\de_z+m_\pi^2e^{\imag\phi})\pi^+,
\label{pipluseom}
\end{equation}
where the prime denotes a derivative with respect to $z$.

Let us first discuss the chiral limit. Solving eq.~\eqref{pipluseom} is now equivalent to the usual Landau level problem. Using eq.~\eqref{CSL_chiral_limit}, the resulting charged pion dispersion relation is given by
\begin{equation}
\omega^2=p_z^2-\frac{\mu Bp_z}{2\pi^2f_\pi^2}+(2n+1)B,
\label{piplusdisp}
\end{equation}
where the non-negative integer $n$ labels the Landau levels. We observe that the bottom of the lowest Landau level reaches zero at a critical magnetic field given by
\begin{equation}
B_\text{BEC}=\frac{16\pi^4f_\pi^4}{\mu^2}.
\label{B_BEC_chiral}
\end{equation}
In yet stronger magnetic fields, charged pions will undergo Bose-Einstein condensation (BEC). To get a rough estimate, we insert the physical value $f_\pi\approx92\text{ MeV}$ and $\mu=900\text{ MeV}$, that is, just below the threshold for nuclear matter formation. For these values, $B_\text{BEC}\approx0.14\text{ GeV}^2$, which is large but possibly achievable in nature.\footnote{Note that the minimum of the dispersion relation~\eqref{piplusdisp} at this magnetic field occurs at $p_z\approx370\text{ MeV}$, which still lies within the range of validity of the chiral perturbation theory, since $4\pi f_\pi\approx1160\text{ MeV}$.} Indeed, the magnetic fields in the cores of magnetars are expected to be roughly in the range $10^{16}$--$10^{19}\text{ G}$~\cite{Andersen:2014xxa}. In our units, $1\text{ GeV}^2\approx1.7\times10^{20}\text{ G}$.

Getting back to the general case, note that the $x$ and $z$ variables are separated in eq.~\eqref{pipluseom}, the former leading to the same Landau level problem. Upon a field redefinition $\pi^+\equiv e^{\imag\phi}\tilde\pi^+$ and using eq.~\eqref{CSL_solution}, the equation of motion~\eqref{pipluseom} can be cast as
\begin{equation}
\left[-\de_{\bar z}^2+6k^2\sn^2(\bar z,k)\right]\tilde\pi^+=\frac{k^2}{m_\pi^2}[\omega^2-(2n+1)B]\tilde\pi^++(k^2+4)\tilde\pi^+.
\end{equation}
The left-hand side of this equation is the Lam\'e operator with $n=2$. According to refs.~\cite{Lame:Finkel,Lame:Iachello,Kishine:2016kis}, its spectrum consists of three bands. The lower edge of the first band corresponds to the eigenvalue $2(1+k^2-\sqrt{1-k^2+k^4})$. Hence, the bottom of the lowest Landau level is given, for arbitrary $m_\pi$, by
\begin{equation}
\min\omega^2_{n=0}=B-\frac{m_\pi^2}{k^2}\left(2-k^2+2\sqrt{1-k^2+k^4}\right),
\label{omegamin}
\end{equation}
where $k$ is given implicitly by eq.~\eqref{condition}.

This calculation confirms the observation made above for the chiral limit, that sufficiently strong magnetic fields lead to an instability of CSL under BEC of charged pions. Indeed, in strong magnetic fields, the elliptic modulus $k$~\eqref{condition} is asymptotically proportional to $1/B$, hence the second term in eq.~\eqref{omegamin} goes as $B^2$. The bottom of the lowest Landau level thus necessarily reaches zero for strong enough $B$.

\begin{figure}
\begin{center}
\includegraphics[width=0.75\textwidth]{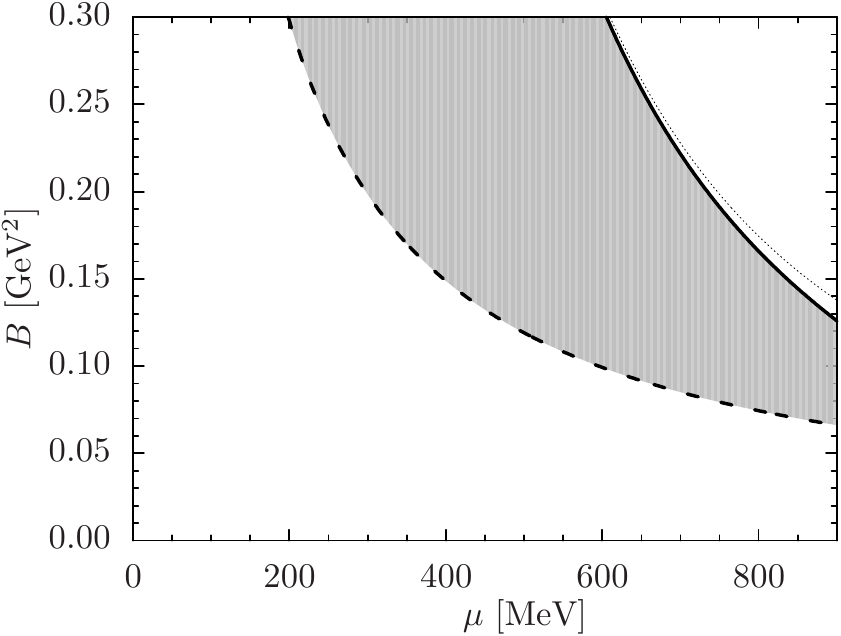}
\end{center}
\caption{The ``phase diagram'' of CSL as a function of chemical potential and magnetic field. CSL becomes favored over the QCD vacuum above the critical field $B_\text{CSL}$ (dashed line). For $B\geq B_\text{BEC}$, BEC of charged pions occurs (solid line). The dotted line stands for $B_\text{BEC}$ in the chiral limit~\eqref{B_BEC_chiral}. The shaded area indicates the region in which CSL of neutral pions \emph{can} exist as the QCD ground state. The physical values $f_\pi\approx92\text{ MeV}$ and $m_\pi\approx140\text{ MeV}$ were used to obtain the numerical results.}
\label{fig:BEC}
\end{figure}

Our main results, eqs.~\eqref{main} and~\eqref{omegamin}, are depicted in figure~\ref{fig:BEC}. We observe that there is a range of magnetic fields in which CSL is energetically favored. However, for higher values of $\mu$, not shown in figure~\ref{fig:BEC}, a direct transition from the QCD vacuum to charged pion BEC might occur. This is due to non-monotonicity of eq.~\eqref{omegamin} as a function of $B$. In addition to being negative for large enough $B$, it also becomes negative at low $B$ above certain threshold $\mu$. The latter coincides with the threshold for stability of a domain wall, which is a degenerate CSL at $B=B_\text{CSL}$. Setting $k=1$, eq.~\eqref{omegamin} reduces to $B-3m_\pi^2$,\footnote{This reproduces the result of ref.~\cite{Son:2007ny} regarding the instability of a single domain wall under charged pion fluctuations.} which together with eq.~\eqref{main} implies instability of the domain wall under charged pion fluctuations at
\begin{equation}
\mu\geq\frac{16\pi f_\pi^2}{3m_\pi}\approx1010\text{ MeV},
\label{mucrit}
\end{equation}
where the physical values of $f_\pi$ and $m_\pi$ were used. However, this result should be regarded as a mere rough estimate; both its numerical value and its inverse proportionality to $m_\pi$ indicate that it may lie outside the range of validity of our effective theory. Unlike the other results obtained above, the prediction of a direct transition from the QCD vacuum to a phase with charged pion BEC should thus be treated with care.

%%%%%%%%%%%%%%%%%%%%%%%%%%%%%%%%%%%%%%%%%%%%%%%%%%%%%%%%%%%%%%%%%%%%

\section{Discussion}

\begin{figure}
\begin{center}
\includegraphics[width=0.75\textwidth]{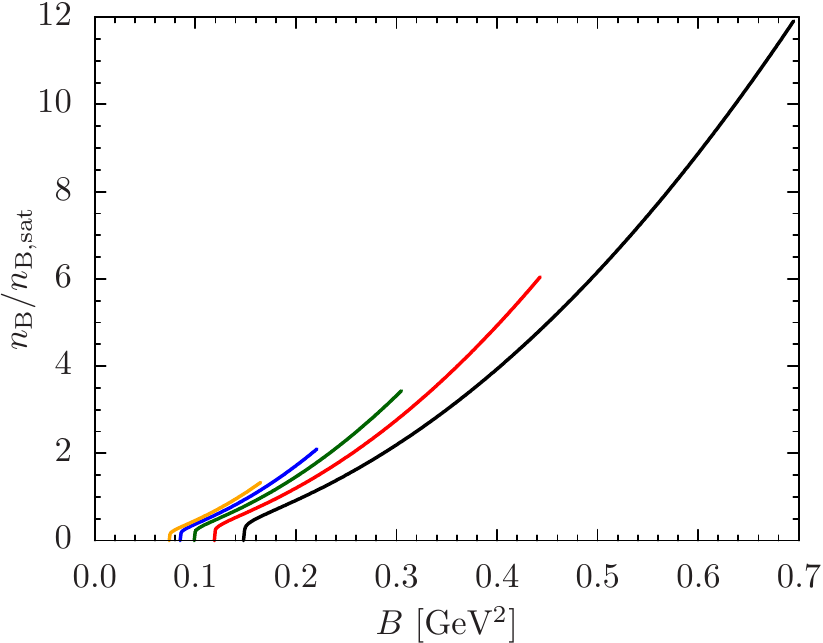}
\end{center}
\caption{Spatially averaged baryon number density carried by the CSL in units of the nuclear saturation density. The individual curves correspond to the values of $\mu$ of $400\text{ MeV}$ (black), $500\text{ MeV}$ (red), $600\text{ MeV}$ (green), $700\text{ MeV}$ (blue), and $800\text{ MeV}$ (orange). For each curve, the onset of nonzero baryon density corresponds to $B_\text{CSL}$, whereas the endpoint indicates the BEC instability at the critical field $B_\text{BEC}$.}
\label{fig:nB}
\end{figure}

We have shown that baryonic matter is unstable under formation of a chiral soliton lattice in the range of magnetic fields, $B_\text{CSL}<B<B_\text{BEC}$. Our results are based on a systematic low-energy effective theory and are model-independent with the relative accuracy of $p_\text{CSL}^2/(4\pi f_\pi)^2$. For $B>B_\text{BEC}$, however, the CSL is itself unstable under BEC of charged pions. Understanding the structure of the ground state in this regime is beyond the scope of the present paper and would be an interesting future problem. The range of magnetic fields in which the CSL appears may be relevant for the physics of magnetars. To demonstrate that the CSL can also reach sufficiently high baryon number densities, relevant in this context, we show in figure~\ref{fig:nB} the spatially averaged density $n_\text{B}$, as given by eq.~\eqref{topcharge}, in units of the nuclear saturation density, $n_\text{B,sat}\approx0.16\text{ fm}^{-3}$.

In order to simplify the discussion, we worked exclusively with the leading-order chiral Lagrangian~\eqref{lagrangian} and only included the effect of the anomaly through the surface term in eq.~\eqref{hamiltonian}. In fact, the anomalous Wess-Zumino-Witten contribution to the Lagrangian adds a term with a time derivative to eq.~\eqref{Lbilin}. However, while this changes somewhat the charged pion dispersion relation~\eqref{piplusdisp}, our result for the critical magnetic field $B_\text{BEC}$, extracted from the condition $\omega_{n=0}=0$, remains valid without modification, see appendix~\ref{app:charged}.

We also simplified the analysis by neglecting the effects of baryon degrees of freedom altogether. It follows from our discussion in section~\ref{sec:ground_state} that in the region of parameter space of interest to us, there are no baryons in the QCD ground state. However, for the highest values of the chemical potential $\mu$, shown in figure~\ref{fig:BEC}, the free energy of nucleons becomes small, and one might therefore wonder if they affect our conclusions through \emph{virtual} corrections. Yet, thanks to the derivative coupling of pions to nucleons, the latter only contribute to pion dynamics at the next-to-leading order of the derivative expansion, and can therefore be consistently omitted from our setup.

Some nonuniform states with one-dimensional modulation are known to be unstable under transverse fluctuations at any nonzero temperature~\cite{Hidaka:2015xza,Lee:2015bva}. Here we assumed zero temperature throughout the text, yet we do not expect such an instability to appear even at nonzero temperature in the situation considered here. Namely, the instability arises from a quadratic dispersion relation of the transverse fluctuations and in turn from rotational invariance of the system. In our setup, the dispersion relation of low-energy excitations is linear, see eq.~\eqref{phonondisp}, and rotational invariance is broken by the magnetic field. The linear dispersion relation in turn justifies the usual power counting rules which dictate that the effects of fluctuations are suppressed by the loop factor $p_\text{CSL}^2/(4\pi f_\pi)^2$~\cite{Weinberg:1978kz,Manohar:1983md}.

Finally, we note that our analysis applies to QCD in an \emph{external} magnetic field. In presence of a dynamical electromagnetic field, the situation changes in several aspects. On the one hand, the anomalous coupling of neutral pions to the electromagnetic field leads to a modification of the photon dispersion relation; in the chiral limit, it exhibits a nonrelativistic behavior~\cite{Yamamoto:2015maz}. On the other hand, the requirement of charge neutrality may \emph{require} that BEC of charged pions localized to the nodes of the soliton lattice accompanies the CSL state~\cite{Son:2007ny}.

%%%%%%%%%%%%%%%%%%%%%%%%%%%%%%%%%%%%%%%%%%%%%%%%%%%%%%%%%%%%%%%%%%%%

\acknowledgments
This work was supported in part by JSPS KAKENHI Grant No.~16K17703 and MEXT-Supported Program for the Strategic Research Foundation at Private Universities, ``Topological Science'' (Grant No.~S1511006). Furthermore, we acknowledge support from the University of Stavanger provided through the conference on Strong and ElectroWeak Matter 2016, where this collaboration was initiated.

%%%%%%%%%%%%%%%%%%%%%%%%%%%%%%%%%%%%%%%%%%%%%%%%%%%%%%%%%%%%%%%%%%%%

\appendix

%%%%%%%%%%%%%%%%%%%%%%%%%%%%%%%%%%%%%%%%%%%%%%%%%%%%%%%%%%%%%%%%%%%%

\section{Effective Lagrangian}
\label{app:EFT}

In this and the following appendices, we provide some details of the calculations, and show how they are modified by the Wess-Zumino-Witten (WZW) contribution to the chiral Lagrangian.

The normal, invariant part of the chiral Lagrangian is given by eq.~\eqref{lagrangian},
\begin{equation}
\La=\frac{f_\pi^2}4\left[\tr(D_\mu\he\Sigma D^\mu\Sigma)+2m_\pi^2\re\tr\Sigma\right],
\label{app:lagrangian}
\end{equation}
where $D_\mu\Sigma\equiv\de_\mu\Sigma-\imag[Q_\mu,\Sigma]$ is the covariant derivative and $Q_\mu\equiv A_\mu\frac{\tau_3}2$. The Chiral Soliton Lattice (CSL) structure is characterized by a neutral pion condensate, corresponding to
\begin{equation}
\ave\Sigma\equiv\Sigma_0=e^{\imag\tau_3\phi}.
\label{app:vacuumansatz}
\end{equation}
It proves convenient to undo the ground state modulation as described by eq.~\eqref{app:vacuumansatz} by redefining the field variable $\Sigma$ as
\begin{equation}
\Sigma\equiv\Sigma_0U,
\end{equation}
where $U$ is a new unitary matrix field such that $\langle U\rangle=\openone$. Inserting all the definitions into the Lagrangian~\eqref{app:lagrangian} and throwing away terms that do not depend on $U$, we get
\begin{equation}
\begin{split}
\La=\frac{f_\pi^2}4\tr\biggl[&\de_\mu\he U\de^\mu U+\imag\tau_3\de^\mu\phi(U\de_\mu\he U-\de_\mu U\he U)+\imag A^\mu\tau_3(\de_\mu U\he U-\he U\de_\mu U)\\
&-\frac12A_\mu A^\mu\tau_3U\tau_3\he U+m_\pi^2(\Sigma_0U+\he U\he\Sigma_0)\biggr].
\end{split}
\label{app:lagrangian2}
\end{equation}
Note that the intrinsic parity symmetry of the Lagrangian~\eqref{app:lagrangian} can be implemented as the replacement $\Sigma\to\he\Sigma$. In terms of the new field variable $U$, this is equivalent to
\begin{equation}
\Sigma_0\to\he\Sigma_0,\qquad
U\to\Sigma_0\he U\he\Sigma_0.
\end{equation}
In case of a coordinate-dependent background $\phi$, the individual terms in the Lagrangian~\eqref{app:lagrangian2} are not invariant under this transformation. However, the Lagrangian as a whole is.

To move on, we next need to choose a parameterization for the unitary matrix $U$. Rather than the standard exponential parameterization, we will choose
\begin{equation}
U(\vec\pi)=\sqrt{1-\frac{\vec\pi^2}{f_\pi^2}}+\frac{\imag\vec\tau\cdot\vec\pi}{f_\pi}.
\label{app:param}
\end{equation}
A short calculation then leads to
\begin{equation}
\begin{split}
\La={}&\frac12\frac{(\de_\mu\vec\pi)^2-\frac1{f_\pi^2}(\vec\pi\times\de_\mu\vec\pi)^2}{1-\frac{\vec\pi^2}{f_\pi^2}}-2f_\pi(\de^\mu\phi)\pi_3\de_\mu\sqrt{1-\frac{\vec\pi^2}{f_\pi^2}}\\
&+(A^\mu-\de^\mu\phi)(\pi_1\de_\mu\pi_2-\pi_2\de_\mu\pi_1)+\frac12A^\mu A_\mu(\pi_1^2+\pi_2^2)\\
&+m_\pi^2f_\pi^2\left(\sqrt{1-\frac{\vec\pi^2}{f_\pi^2}}-1\right)\left(\cos\phi+\frac{\pi_3}{f_\pi}\sin\phi\right)
\end{split}
\end{equation}
plus a surface term. For the record, we also put down explicitly the bilinear part of this Lagrangian, eq.~\eqref{Lbilin} of the main text,
\begin{equation}
\La_\text{bilin}=\frac12(\de_\mu\vec\pi)^2+(A^\mu-\de^\mu\phi)(\pi_1\de_\mu\pi_2-\pi_2\de_\mu\pi_1)+\frac12 A^\mu A_\mu(\pi_1^2+\pi_2^2)-\frac12m_\pi^2\vec\pi^2\cos\phi.
\label{app:Lbilin}
\end{equation}
We assumed that the gauge for $A^\mu$ has been chosen so that $A^\mu\de_\mu\phi=0$, which is always possible in a uniform magnetic field.

The anomalous, WZW contribution to the action reads~\cite{Son:2007ny}
\begin{equation}
S_\text{WZW}=-\int\dd^4x\left(A_\mu^\text{B}-\frac12A_\mu\right)j^\mu_\text{B},
\end{equation}
where $A^\text{B}_\mu$ is an external gauge field that couples to the baryon number and $j^\mu_\text{B}$ is the Goldstone-Wilczek current. The latter is given by the manifestly covariant expression
\begin{equation}
j^\mu_\text{B}=-\frac1{24\pi^2}\epsilon^{\mu\nu\alpha\beta}\tr\biggl[(\Sigma D_\nu\he\Sigma)(\Sigma D_\alpha\he\Sigma)(\Sigma D_\beta\he\Sigma)+\frac{3\imag}4F_{\nu\alpha}\tau_3(\Sigma D_\beta\he\Sigma+D_\beta\he\Sigma\Sigma)\biggr].
\end{equation}
Upon inserting the definition of the covariant derivative, the current acquires the form
\begin{equation}
\begin{split}
j^\mu_\text{B}=-\frac1{24\pi^2}\epsilon^{\mu\nu\alpha\beta}\tr\biggl[&-\Sigma\de_\nu\he\Sigma\de_\alpha\Sigma\de_\beta\he\Sigma+3\imag Q_\nu(\de_\alpha\he\Sigma\de_\beta\Sigma-\de_\alpha\Sigma\de_\beta\he\Sigma)\\
&+\frac{3\imag}4F_{\nu\alpha}\tau_3(\Sigma\de_\beta\he\Sigma+\de_\beta\he\Sigma\Sigma)\biggr].
\end{split}
\end{equation}
Once the field fluctuations are separated from the ground state by defining $\Sigma=\Sigma_0U$ as above, the current can by a longer manipulation be cast as
\begin{equation}
j^\mu_\text{B}=\frac1{24\pi^2}\epsilon^{\mu\nu\alpha\beta}\tr(U\de_\nu\he U\de_\alpha U\de_\beta\he U)+\de_\nu G^{\mu\nu},
\end{equation}
where
\begin{equation}
\begin{split}
G^{\mu\nu}=-\frac1{16\pi^2}\epsilon^{\mu\nu\alpha\beta}\Bigl\{&\phi F_{\alpha\beta}+2\imag\phi\tr(\tau_3\de_\alpha U\de_\beta\he U)+A_\alpha\de_\beta\phi\tr(\tau_3\he U\tau_3U)\\
&+\imag A_\alpha\tr[\tau_3(U\de_\beta\he U+\de_\beta\he UU)]\Bigr\}.
\end{split}
\end{equation}
The first term in $G^{\mu\nu}$ is independent of the pion fields. Together with the pion-independent piece of the third term (obtained by the replacement $U\to\openone$), this is responsible for the stabilization of the CSL solution in an external magnetic field. It is easy to see that in a uniform external magnetic field, only the second term can contribute to perturbative pion physics. Using the parameterization~\eqref{app:param}, we can work out the part of the Lagrangian bilinear in the pion fields explicitly. Including the topological term, it reads
\begin{equation}
\La_\text{WZW}^\text{bilin}=\frac1{8\pi^2}\epsilon^{\mu\nu\alpha\beta}A^\text{B}_\mu\de_\nu(\phi F_{\alpha\beta})-\frac1{16\pi^2f_\pi^2}\epsilon^{\mu\nu\alpha\beta}F_{\mu\nu}\de_\alpha\phi(\pi_1\de_\beta\pi_2-\pi_2\de_\beta\pi_1).
\label{app:LbilinWZW}
\end{equation}

%%%%%%%%%%%%%%%%%%%%%%%%%%%%%%%%%%%%%%%%%%%%%%%%%%%%%%%%%%%%%%%%%%%%

\section{Neutral pion fluctuations}
\label{app:phonon}

Thanks to the fact that the third component of the isospin $\text{SU}(2)$ symmetry remains intact by the neutral pion background, the neutral and charged pion fluctuations do not mix and can be treated separately. Since the WZW term~\eqref{app:LbilinWZW} obviously includes only the charged pions, the discussion of the phonon spectrum in the main text is not affected by it. Here we wish to give an alternative derivation of the phonon group velocity, eq.~\eqref{phonondisprel} in the main text, which makes use of the explicit Bloch-type wave function of the phonon.

The Lam\'e equation takes the general form
\begin{equation}
[-\de_{\bar z}^2+n(n+1)k^2\sn^2(\bar z,k)]\psi(\bar z)=A\psi(\bar z).
\label{app:lame}
\end{equation}
For $n=1$, its solution is known to be of the form~\cite{Whittaker:1927,Sutherland:1973zz}
\begin{equation}
\psi(\bar z)=\frac{H(\bar z+\sigma,k)}{\Theta(\bar z,k)}e^{-\bar z Z(\sigma,k)},
\label{app:solution}
\end{equation}
where $H$, $\Theta$, and $Z$ denote Jacobi's eta, theta, and zeta functions and $\sigma$ is a parameter which is related to $A$ by
\begin{equation}
A=1+k^2\cn^2(\sigma,k).
\label{app:dispersion}
\end{equation}
The solution~\eqref{app:solution} can be rewritten equivalently as
\begin{equation}
\psi(\bar z)=\frac{H(\bar z+\sigma,k)}{\Theta(\bar z,k)}\exp\left[-\imag\pi\frac{\bar z}{2K(k)}\right]\exp\left\{\imag\bar z\left[\imag Z(\sigma,k)+\frac\pi{2K(k)}\right]\right\}.
\end{equation}
This is the usual Bloch form of the wave function. The first two terms together are explicitly periodic in $\bar z$ with period $2K(k)$, and hence in $z$ with period $2kK(k)/m_\pi=\ell$. The last term takes the form $e^{\imag p_zz}=e^{\imag k\bar zp_z/m_\pi}$, which leads to the identification of the crystal momentum as
\begin{equation}
p_z=\frac{m_\pi}k\left[\imag Z(\sigma,k)+\frac\pi{2K(k)}\right]=\frac{\imag m_\pi}kZ(\sigma,k)+\frac\pi\ell.
\label{app:quasimomentum}
\end{equation}
Obviously, the solution is only bounded (and thus physically acceptable) when $Z(\sigma,k)$ is purely imaginary. Within the period of $Z$, this happens only for $\re\sigma$ equal to $0$ or $K(k)$. The spectrum of solutions to the Lam\' e equation~\eqref{app:lame} with $n=1$ therefore consists of two bands, corresponding to
\begin{equation}
\sigma_v=K(k)+\imag\kappa,\qquad
\sigma_c=\imag\kappa,
\end{equation}
where $\kappa$ is a real parameter. Here $\sigma_v$ corresponds to the ``valence band'' that encodes the phonon excitation, whereas $\sigma_c$ corresponds to the ``conduction band''.

We would now like to find a closed analytic expression for the phonon dispersion relation. To that end, we first note that the bottom of the valence band appears at $\kappa=K(k')$, where $k'\equiv\sqrt{1-k^2}$ is the complementary modulus. Shifting the parameter $\kappa$ to the minimum by $\delta\kappa\equiv\kappa-K(k')$ and using the periodicity properties of the Jacobi elliptic functions, one can prove that
\begin{equation}
\cn(\sigma_v,k)=\cn(K(k)+\imag K(k')+\imag\delta\kappa,k)=-\frac{\imag k'}k\cn(\delta\kappa,k').
\end{equation}
Hence for the valence band,
\begin{equation}
A=1-k'^2\cn^2(\delta\kappa,k'),
\end{equation}
and using eq.~\eqref{eomaux} of the main text, we can identify the dispersion relation of the phonon. Namely, for a purely longitudinal motion, $p_x=p_y=0$, we get
\begin{equation}
\frac\omega{m_\pi}=\frac{k'}k\sn(\delta\kappa,k')\approx\frac{k'}k\delta\kappa+\mathcal O(\delta\kappa^3).
\label{app:dispaux}
\end{equation}

It remains to find an explicit relation between the parameter $\kappa$ and the crystal momentum $p_z$ at the bottom of the valence band. To that end, we use the identity
\begin{equation}
\frac{\dd Z(\sigma,k)}{\dd\sigma}=\dn^2(\sigma,k)-\frac{E(k)}{K(k)}.
\label{app:dZu}
\end{equation}
As above, the periodicity properties of the Jacobi elliptic functions tell us that 
\begin{equation}
\dn(\sigma_v,k)=\dn(K(k)+\imag K(k')+\imag\delta\kappa,k)=-k'\sn(\delta\kappa,k'),
\end{equation}
which manifestly vanishes at $\delta\kappa=0$. We thus deduce from eq.~\eqref{app:dZu} and from the definition of Bloch momentum~\eqref{app:quasimomentum} that
\begin{equation}
\frac{\dd p_z}{\dd\kappa}=\frac{m_\pi E(k)}{kK(k)}
\end{equation}
at the bottom of the valence band. Together with eq.~\eqref{app:dispaux}, this then immediately gives the desired expression for the phonon group velocity,
\begin{equation}
c_\text{ph}=\sqrt{1-k^2}\frac{K(k)}{E(k)},
\end{equation}
in accord with eq.~\eqref{phonondisprel} in the main text.

%%%%%%%%%%%%%%%%%%%%%%%%%%%%%%%%%%%%%%%%%%%%%%%%%%%%%%%%%%%%%%%%%%%%

\section{Charged pion fluctuations}
\label{app:charged}

Since the WZW term~\eqref{app:LbilinWZW} carries one time derivative of the charged pion fields, it does not affect the discussion of the instability with respect to charged pion Bose-Einstein condensation (BEC),  which is defined by the condition that the bottom of the lowest Landau level appears at $\omega=0$. Below, we review the modifications that the WZW term does bring.

In order to extract the spectrum of the charged pion fluctuations, we need both eqs.~\eqref{app:Lbilin} and~\eqref{app:LbilinWZW}. As in the main text, we choose the gauge for the vector potential as $\vek A=(0,Bx,0)$, and introduce the charge eigenstates $\pi^\pm$ via
\begin{equation}
\pi^\pm\equiv\frac1{\sqrt2}(\pi_1\pm\imag\pi_2).
\end{equation}
Upon Fourier transforming in time and the $y$-coordinate, the equation of motion for $\pi^+$, following from the bilinear Lagrangian, takes the two-dimensional form
\begin{equation}
\omega^2\pi^+=\left[-\de_x^2+B^2\left(x-\frac{p_y}B\right)^2\right]\pi^++\left(-\de_z^2+2\imag\phi'\de_z+m_\pi^2e^{\imag\phi}-\frac{\omega B\phi'}{4\pi^2f_\pi^2}\right)\pi^+,
\end{equation}
which generalizes eq.~\eqref{pipluseom} of the main text. The right-hand side of this equation consists of two mutually commuting operators, whose spectrum can therefore be determined separately. The first of these represents the standard Landau level problem, and we can thus write the spectrum right away as
\begin{equation}
\omega^2=(2n+1)B+\lambda,
\label{app:omegalambda}
\end{equation}
where $\lambda$ runs over the eigenvalues of the one-dimensional differential operator
\begin{equation}
\Delta\equiv-\de_z^2+2\imag\phi'\de_z+m_\pi^2e^{\imag\phi}-\frac{\omega B\phi'}{4\pi^2f_\pi^2}.
\label{app:Delta}
\end{equation}

In the chiral limit, the ground state configuration is given by a linear function of $z$, eq.~\eqref{CSL_chiral_limit} of the main text. The operator~$\Delta$ is then diagonalized by a further Fourier transform in the $z$-coordinate,
\begin{equation}
\lambda=p_z^2-\frac{\mu Bp_z}{2\pi^2f_\pi^2}-\frac{\mu\omega B^2}{16\pi^4f_\pi^4}.
\end{equation}
The dispersion relation of the charged pions then immediately follows from eq.~\eqref{app:omegalambda},
\begin{equation}
\omega=-\frac{\mu B^2}{32\pi^4f_\pi^4}+\sqrt{\left(\frac{\mu B^2}{32\pi^4f_\pi^4}\right)^2+p_z^2-\frac{\mu Bp_z}{2\pi^2f_\pi^2}+(2n+1)B}.
\end{equation}
This modifies the dispersion relation found in the main text, eq.~\eqref{piplusdisp}. However, it does not change the conclusion that the bottom of the lowest Landau level drops to zero at
\begin{equation}
B=B_\text{BEC}\equiv\frac{16\pi^4f_\pi^4}{\mu^2}.
\end{equation}

Away from the chiral limit, the spectrum of the operator~$\Delta$ given by eq.~\eqref{app:Delta} does not seem to be easy to find. Similarly to the main text, we can simplify it by a unitary transformation $\Delta\to\tilde\Delta\equiv e^{-\imag\phi}\Delta e^{\imag\phi}$, which upon using the equation of motion for the background $\phi$ leads to
\begin{equation}
\tilde\Delta=-\de_z^2-(\phi')^2+m_\pi^2\cos\phi-\frac{\omega B\phi'}{4\pi^2f_\pi^2}.
\end{equation}
However, the presence of the last, anomalous term prevents its straightforward reduction to the Lam\'e operator. Fortunately, this last term vanishes at $\omega=0$, and thus the conclusion made in the main text regarding the critical magnetic field $B_\text{BEC}$ for BEC of charged pions still holds.

%%%%%%%%%%%%%%%%%%%%%%%%%%%%%%%%%%%%%%%%%%%%%%%%%%%%%%%%%%%%%%%%%%%%

\bibliographystyle{JHEP}
\bibliography{references}

\end{document}